\documentclass[11pt,preprintnumbers,aps,amssymb,nofootinbib,amsmath]
{revtex4}
\usepackage{epsfig,epsf}
\usepackage{bm} 
\newcommand{\beq}{\begin{equation}}
\newcommand{\beql}[1]{\begin{equation}\label{#1}}
\newcommand{\eeq}{\end{equation}}
\newcommand{\bea}{\begin{eqnarray}}
\newcommand{\eea}{\end{eqnarray}}
\newcommand{\eq}[1]{(\ref{#1})}
\newcommand{\fig}[1]{Fig.~\ref{#1}}
\renewcommand{\sec}[1]{Sec.~\ref{#1}}
\newcounter{topiccounter}
\renewcommand{\v}[1]{{\vec {#1}}}
\newcommand{\unit}[1]{\hat {\mathbf{#1}}} 

\newcommand{\as}{\alpha_s}

\newcommand{\e}{\varepsilon}

\begin{document}

\preprint{RBRC-845}

\title{Synchrotron radiation  by fast fermions  in heavy-ion collisions}

\author{Kirill Tuchin$\,^{a,b}$\\}

\affiliation{
$^a\,$Department of Physics and Astronomy, Iowa State University, Ames, IA 50011\\
$^b\,$RIKEN BNL Research Center, Upton, NY 11973-5000\\}

\date{\today}

\pacs{}

\begin{abstract}
We study the synchrotron radiation of gluons by fast quarks in strong magnetic field produced by colliding relativistic heavy-ions. We argue that  due to high electric conductivity of plasma, time variation of the magnetic field is very slow and estimate its relaxation time.  We calculate the energy loss due to synchrotron radiation of gluons by fast quarks.  We find that the typical energy loss per unit length for a light quark at LHC is a few GeV per fm.  This effect alone predicts quenching of jets with $p_\bot$ up to about 20 GeV. We also show that the spin-flip transition effect accompanying the synchrotron radiation leads  to a strong polarization of quarks and leptons with respect to the direction of the magnetic field.  Observation of the lepton polarization may provide a direct evidence of existence of strong magnetic field in heavy-ion collisions. 

\end{abstract}

\maketitle

\section{Introduction}\label{sec:intr}

It has been recently argued \cite{Kharzeev:2007jp} that the magnetic field $B$ produced by colliding relativistic heavy-ions can be as large as $eB\approx m_\pi^2/\hbar$ at RHIC and $eB\approx 15\,m_\pi^2/\hbar$ at LHC \cite{Kharzeev:2007jp,Skokov:2009qp}.  This is comparable to the Schwinger critical value $eB_c = m^2/\hbar$ for a quark of mass $m$. In such strong fields many interesting perturbative and  non-perturbative phenomena  are expected to take place (see e.g.\ \cite{Kharzeev:2004ey,Kharzeev:2007tn,Fukushima:2008xe,Kharzeev:2007jp,Basar:2010zd,Asakawa:2010bu}). In this letter we discuss the energy loss and polarization of fast light quarks moving in external magnetic field \cite{Nikishov:1964zza,Sokolov:1963zn}.  In QED these  phenomena have been studied in detail due to their significance for collider physics, see e.g.\ \cite{Berestetsky:1982aq,Jackson:1975qi}. In turn, energy loss of fast particles in heavy-ion collisions is one of the most important probes of the hot nuclear medium \cite{Gyulassy:1993hr,Baier:1994bd}. Synchrotron radiation in chromo-magnetic fields was discussed in \cite{Shuryak:2002ai,Kharzeev:2008qr,Zakharov:2008uk}.

\begin{figure}[ht]
      \includegraphics[height=4cm]{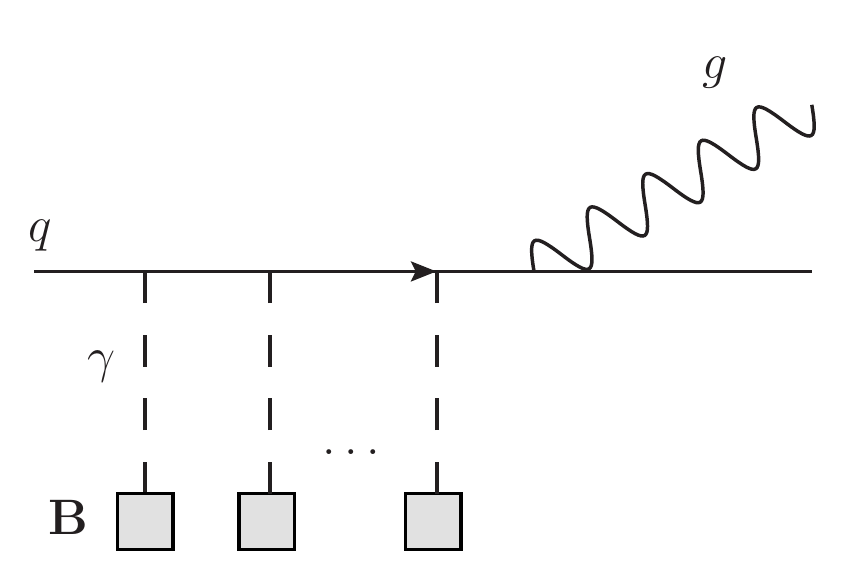} 
  \caption{A typical diagram contributing to the synchrotron radiation by a quark. }
\label{bremss}
\end{figure}

A typical diagram contributing to the synchrotron radiation, i.e.\ radiation in external magnetic field, by a quark is shown in \fig{bremss}. This diagram is proportional to $(eB)^n$, where $n$ is the number of external field lines. In strong field,  powers of $eB$ must be summed up,  which may be accomplished by exactly solving the Dirac equation for  the relativistic fermion and then calculating the matrix element for the transition $q\to q+g$. 
Such calculation has been done in QED for some special cases including the  homogeneous constant field and can be readily generalized for gluon radiation. 

\section{Time dependence of magnetic field }\label{sec:time}

At first we would like to determine whether the constant field approximation is  applicable to magnetic fields created in relativistic heavy-ion collisions. 
For periodic fields, the measure of how fast the field varies is the Keldysh ``adiabaticity" parameter $\gamma$ defined as \cite{Keldysh,Ritus-dissertation}
\beql{gamma}
\gamma = \frac{m\omega_B}{eB_m}\,,
\eeq
where $\omega_B $ is frequency and $B_m$ is amplitude of the magnetic field. We can adopt  $\gamma$ for adiabatically changing fields as well. In this case  $  \omega_B =|\dot B/ B|$ is the rate of the field change and $B_m$ 
is average field over time $1/\omega$. The constant field approximation corresponds to $\gamma\ll 1$. 

Let us now estimate $\gamma$. At first, we assume that $B$ is determined only by the valence charges of initial nuclei and neglect the magnetic response of the produced nuclear matter.  In this case $B$ decreases with time $t$ according to the power law implying that  $\omega_B\sim 1/t$ \cite{Kharzeev:2007jp}, where $t$ is the time measured in the center-of-mass frame. Obviously, at early times  $\omega_B$ is largest. Therefore, to set the upper bound on $\gamma$ we need to estimate the time $t_0$ after a heavy-ion collision  when the quarks are released from the nuclear wave functions. If $Q_s$ is the saturation momentum, this time is $t_0\sim 1/Q_s$. For a semi-peripheral collisions of Gold nuclei at RHIC we obtain $\omega_B=Q_s\approx 1$~GeV. 
 Thus,  the adiabaticity for $u$-quark is $\gamma = 0.1
-0.2$, while for $d$-quark $\gamma= 0.5-0.8$ which marginally justifies the constant field approximation. Here we used  $m_u=1.5-3.3$~MeV and $m_d=3.5-6$~MeV \cite{Amsler:2008zzb}. For heavier quarks $\gamma\gg 1$ and it seems that we cannot apply the constant field approximation. 
Energy dependence of the magnetic field follows from its transformation properties under boosts $eB\propto e^Y$, where $Y$ is the rapidity. On the other hand, $Q_s\propto e^{\lambda Y/2}$, with $\lambda\ll1$ \cite{Gribov:1984tu,GolecBiernat:1999qd}. Therefore, we expect that $\gamma$ will decrease with energy improving the applicability of the constant field approximation. 

So far we have neglected the magnetic response of the quark-gluon plasma. The plasma seems to be strongly coupled and we expect that its dynamics in strong external magnetic field is highly non-trivial. Solution of the problem of plasma magnetic response requires extensive numerical simulations of the relativistic hydrodynamic equations.  Bearing this in mind, we, however, can use semi-classical arguments to derive a simple estimate of how the time dependence is affected by the plasma magnetic response.  We will denote the magnetic field due to the valence quarks as $\vec B_0$.  
According to the Faraday law, decreasing  magnetic field $\vec B_0$ induces electric field $\vec E$ circulating around the direction of magnetic field. The electric field generates electric current  that in turn  produces the magnetic field $\vec B_i$ pointing in the positive $z$ direction according to the Lenz rule. In the adiabatic approximation the resulting total magnetic field $\vec B$ satisfies the following equations
 \cite{LL8}
\beql{m1}
\nabla^2 \vec B = \mu\sigma\,\partial_t{\v B}\,,\quad \v\nabla\cdot \v B=0\,.
\eeq
Here  $\mu$ and $\sigma$ are electric permeability and  conductivity of plasma. The initial condition at $t=t_0$ reads 
\beql{m2}
\vec B(\vec r, t_0)=  \vec B_m\, e^{-\frac{\rho^2}{R^2}}\,,
\eeq
where we used the cylindrical coordinates $\{z,\rho, \phi\}$ and $R$ is of the order of nuclear radius. We neglect the external magnetic field at $t>t_0$. 

Solution to the problem \eq{m1},\eq{m2} is
\beql{m4}
\vec B(\vec r, t) =\int dV' \vec B(\vec r',t_0)\,G(\vec r-\vec r', t-t_0)\,.
\eeq
where 
\beql{green}
G(\vec r, t)= \frac{1}{(4\pi t/\sigma)^{3/2}}\exp\left[ -\frac{\vec r^2}{4t/\sigma}\right]\,
\eeq
the Green's function  and  we assumed $\mu=1$. Eqs.~\eq{m4},\eq{green} describe evolution of the initial field \eq{m2} in time. Integrating over the entire volume  $V'$ we derive
\beql{b1}
\vec B(\vec r, t) = \vec B_m\,\frac{R^2}{R^2+4(t-t_0)/\sigma}\,\exp\left[ -\frac{\rho^2}{R^2+4(t-t_0)/\sigma}\right]\,.
\eeq

It follows from \eq{b1} that as long as $t-t_0\ll \tau$, where $\tau$ is a characteristic time 
\beql{t-relax}
\tau = \frac{R^2\sigma}{4}\,.
\eeq
the magnetic field $\vec B$ is approximately time-independent.

To estimate $\tau$ we use the value of the electric conductivity found in the lattice calculations $\sigma \approx 7C_{EM}T^2/T_c$ \cite{Gupta:2003zh}, where $C_{EM}= 4\pi\alpha_{EM}\sum_f e^2_f$. We have $\sigma= 0.43\, T_c\,(T/T_c)^2$. For  $T_c= 170$~MeV this becomes $\sigma \approx 0.37$~fm$^{-1}\, (T/T_c)^2$. With a conservative estimate of the medium size  $R\approx 5$~fm we have $\tau \approx 2.2\,\text{fm}\,(T/T_c)^2$. Substituting $\omega_B=1/\tau$ into \eq{gamma} we obtain $\gamma< 0.1$ even for $T=T_c$.  Moreover, recent lattice calculations show that the electric conductivity increases in the presence of strong magnetic field \cite{Buividovich:2010tn}. As plasma expands it cools down;  at $T<T_c$ electric conductivity rapidly decreases and magnetic field vanishes.  We thereby conclude -- keeping in mind the qualitative nature of our  argument -- that the magnetic field can be considered as approximately stationary during the process of synchrotron radiation. 

We could have concluded  that the magnetic field is quasi-stationary by mere examining the diffusion equation \eq{m1}. Indeed, on dimensional grounds magnetic field significantly varies over the time $\tau \sim R^2 \sigma$, cp.\ \eq{t-relax}. This argument is arguably qualitative. A quantitative analysis must include a more realistic contribution of external sources, effect  of plasma expansion and realistic transverse geometry. The contribution of external sources would appear in \eq{m4} as an additional term involving integration over space and time (from $t_0$ to $t$) of $\nabla\times j_\text{e}$, where $ j_\text{e}$ is the current density of valence quarks of receding nuclear remnants. Such corrections will certainly introduce time variations of the magnetic field. However, since $\tau$ is more than an order of magnitude larger than the plasma life-time, we expect that adiabatic approximation  $\gamma\ll 1$ still holds in a more accurate treatment. 
Of course, as has been already pointed out, a comprehensive numerical simulation is required in order to obtain the precise time and space dependance of the magnetic field.

In light of our discussion in this section, we will treat magnetic field as slowly varying in time. Formulas for energy loss  in  \sec{sec:el} should be understood as  time average, and $B$ as a mean magnetic field. On the other hand,  polarization effects discussed in \sec{sec:pol}, are sensitive only to the initial value of the magnetic field $B(t_0)$.

\section{Energy loss due to synchrotron radiation}\label{sec:el}

Quark propagating in the external magnetic field radiates gluons as depicted in \fig{bremss}. The intensity of the radiation can be expressed via the invariant parameter $\chi$ defined as 
\beq\label{chi}
\chi^2 =-\frac{\alpha_\mathrm{em}Z_q^2\hbar^3}{m^6}\,(F_{\mu\nu}p^\nu)^2 = \frac{\alpha_\mathrm{em}Z_q^2\hbar^3}{m^6}(\v p\times \v B)^2
\eeq
where the initial quark 4-momentum is $p^\mu=(\e,\v p)$, $Z_q$ is the quark charge in units of the absolute value of the electron charge $e$. At high energies,
\beql{chi-appr}
\chi \approx \frac{Z_qB\e}{B_cm}\,.
\eeq
The regime of weak fields corresponds to $\chi\ll 1$, while in strong fields $\chi \gg 1$. In our case, $eB/eB_c \approx (m_\pi/m_u)^2\gg 1$ (at RHIC) and therefore $\chi\gg 1$.  
In terms of $\chi$, spectrum of radiated gluons of frequency $\omega$ can be written as \cite{Nikishov:1964zza}
\beq\label{glue-spec}
\frac{dI}{d\omega} = -\as C_F \,\frac{m^2 \,\omega}{\e^2}\left\{ 
\int_x^\infty \mathrm{Ai}(\xi)\,d\xi +
\left( \frac{2}{x}+\frac{\omega}{\e}\,\chi \, x^{1/2}\right)\mathrm{Ai}'(x)
\right\}\,,
\eeq
where $I$ is the intensity,
$$
x= \left( \frac{\hbar\omega}{\e'\chi}\right)^{2/3}\,,
$$
and $\e'$ is the quark's energy in the final state.    $\mathrm{Ai}$ is the Ayri function. Eq.~\eq{glue-spec} is valid under the assumption that the initial quark remains ultra-relativistic, which implies that the energy loss due to the synchrotron radiation $\Delta\e$ should be small compared to the quark energy itself $\Delta \e\ll \e$.

Energy loss by a relativistic quark per unit length  is given by \cite{Berestetsky:1982aq}
\beql{enegyloss}
\frac{d\e}{dl}=- \int_0^\infty d\omega \frac{dI}{d\omega} = \alpha_s C_F\,\frac{m^2\,\chi^2}{2}\,\int_0^\infty\frac{4+5\chi\, x^{3/2}+4\chi^2\,x^3}{(1+\chi\, x^{3/2})^4}\,\mathrm{Ai}'(x)\,x\,dx\,.
\eeq
In two interesting limits, energy loss behaves quite differently. At $\eta=\varphi=0$ we have \cite{Berestetsky:1982aq}
\begin{subequations}
 \begin{eqnarray}\label{est}
\frac{d\e}{dl}&=&-\frac{2\,\alpha_s\,\hbar\, C_F\,(Z_qeB)^2\e^2}{3m^4}\,,\quad \chi\ll 1\,,\\
\frac{d\e}{dl}&=&-0.37\,\as \,\hbar^{-1/3}\,C_F\,\left( Z_qeB\, \e \right)^{2/3}\,,\quad \chi\gg 1\,.
\end{eqnarray}
\end{subequations}
In the strong field limit energy loss is independent of the quark mass, whereas in the weak field case it decreases as $m^{-4}$. 
Since $\chi\propto \hbar$, limit of $\chi\ll 1$ corresponds to the classical energy loss. 

\begin{figure}[ht]
      \includegraphics[height=9cm]{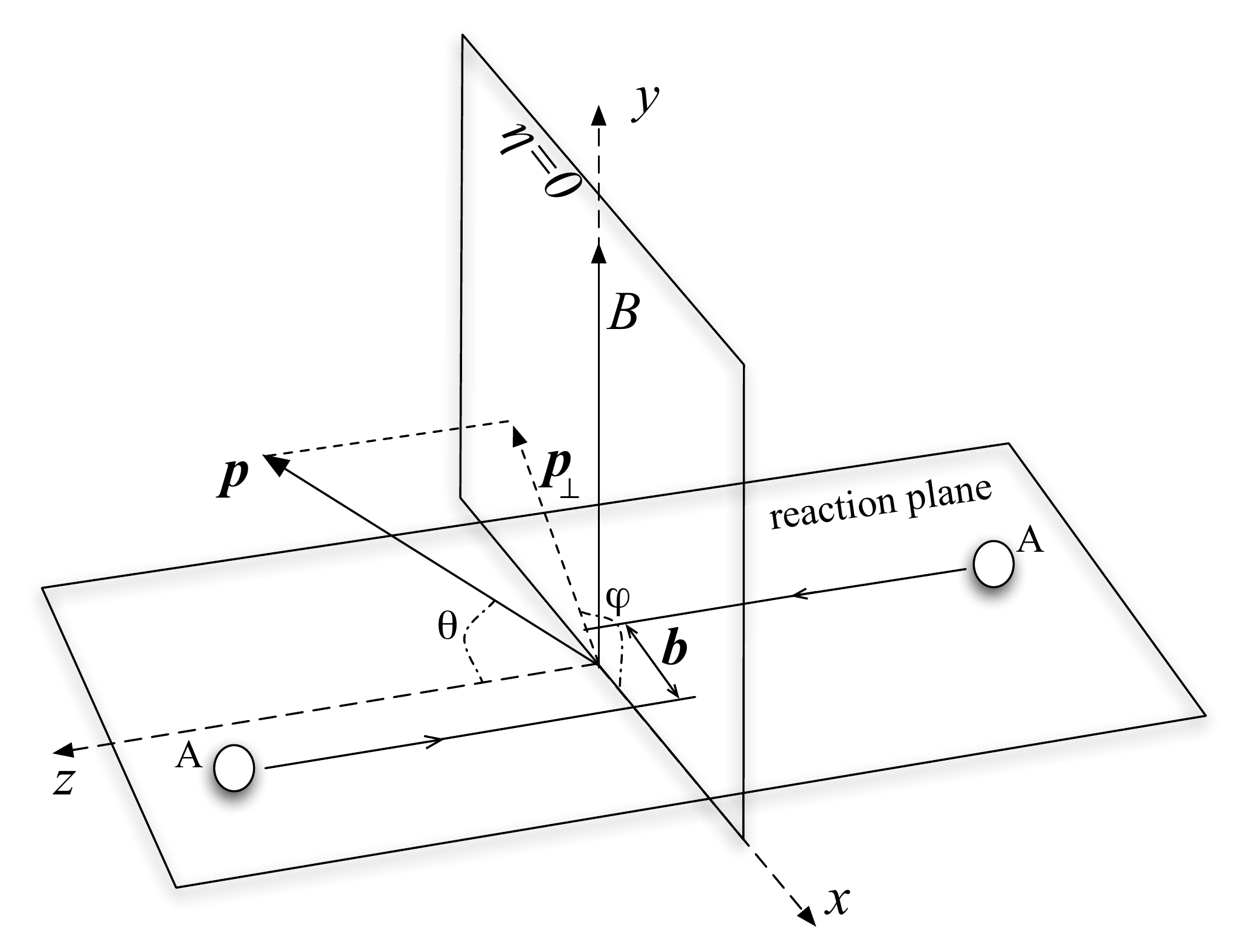} 
  \caption{Geometry of a heavy-ion collision. $\v p$ denotes the quark momentum.  Two orthogonal planes are the ``reaction plane" in which the initial heavy-ion momenta lie, and the mid-rapidity plane $\theta=\pi/2$, which is labeled as $\eta=0$. }
\label{geom}
\end{figure}

To apply this result to heavy-ion collisions we need to write down the invariant $\chi$ in a suitable kinematic variables. The geometry of a heavy-ion collision is depicted in \fig{geom}. We can see that the vector of magnetic field $\vec B$ is orthogonal to the ``reaction plane", which is  spanned by the impact parameter vector $\vec b$ and  the collision axis ($z$-axes). For a quark of momentum $\vec p$ we define the polar angle $\theta$ with respect to the $z$-axis and azimuthal angle $\varphi$ with respect to the reaction plane. 
In this notation, $\v B= B\,\unit y$ and $\v p = p_z\unit z+p_\bot (\unit x\, \cos\varphi+\unit y\sin\varphi)$, where $ p_\bot = |\v p|\sin\theta\approx \e \sin\theta$. Thus, $(\v B\times\v p)^2= B^2(p_z^2+p_\bot^2\cos^2\varphi)$. Conventionally, one expresses the longitudinal momentum and energy using the rapidity $\eta$ as $\e = m_\bot \cosh\eta$ and $p_z= m_\bot \sinh\eta$, where $m_\bot^2= m^2+p_\bot^2$.
We have
\beql{chi-hi}
\chi^2= \frac{\hbar^2(e B)^2}{m^6}\,p_\bot^2 (\sinh^2\eta+\cos^2\varphi)
\eeq


\begin{figure}[ht]
\begin{tabular}{cc}
      \includegraphics[height=1.8in]{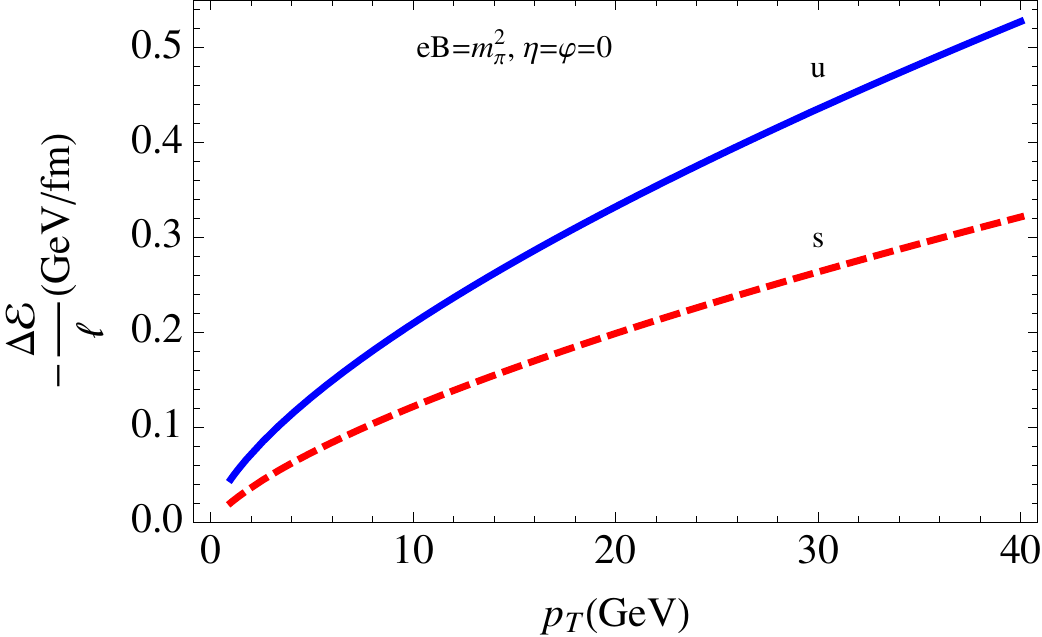} &
      \includegraphics[height=1.8in]{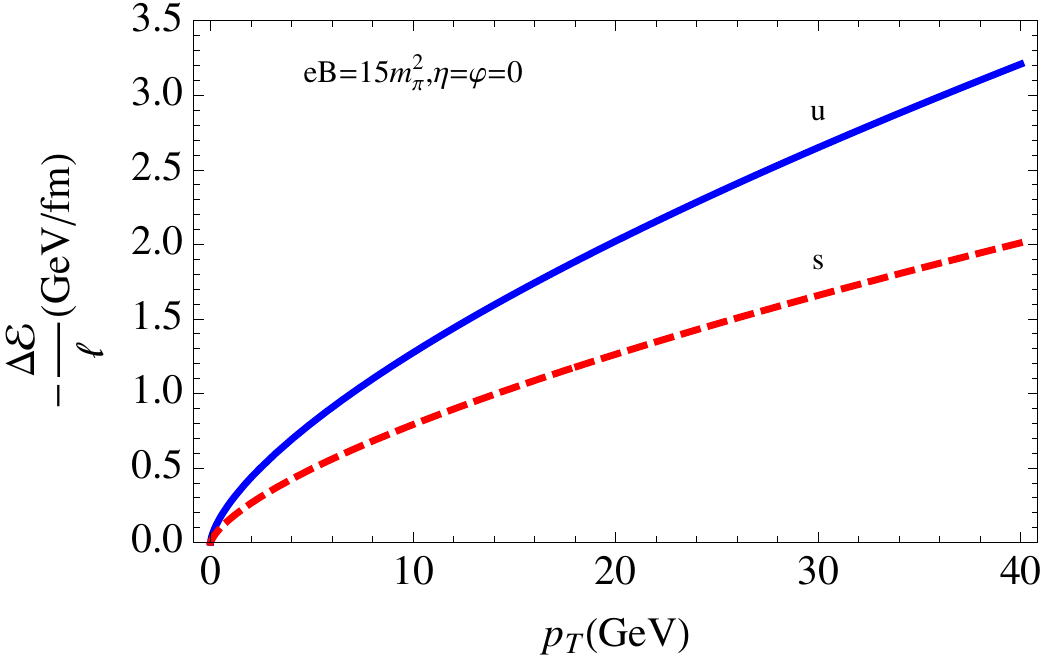}\\
      $(a)$ & $(b)$  \\[0.2in]
        \includegraphics[height=1.8in]{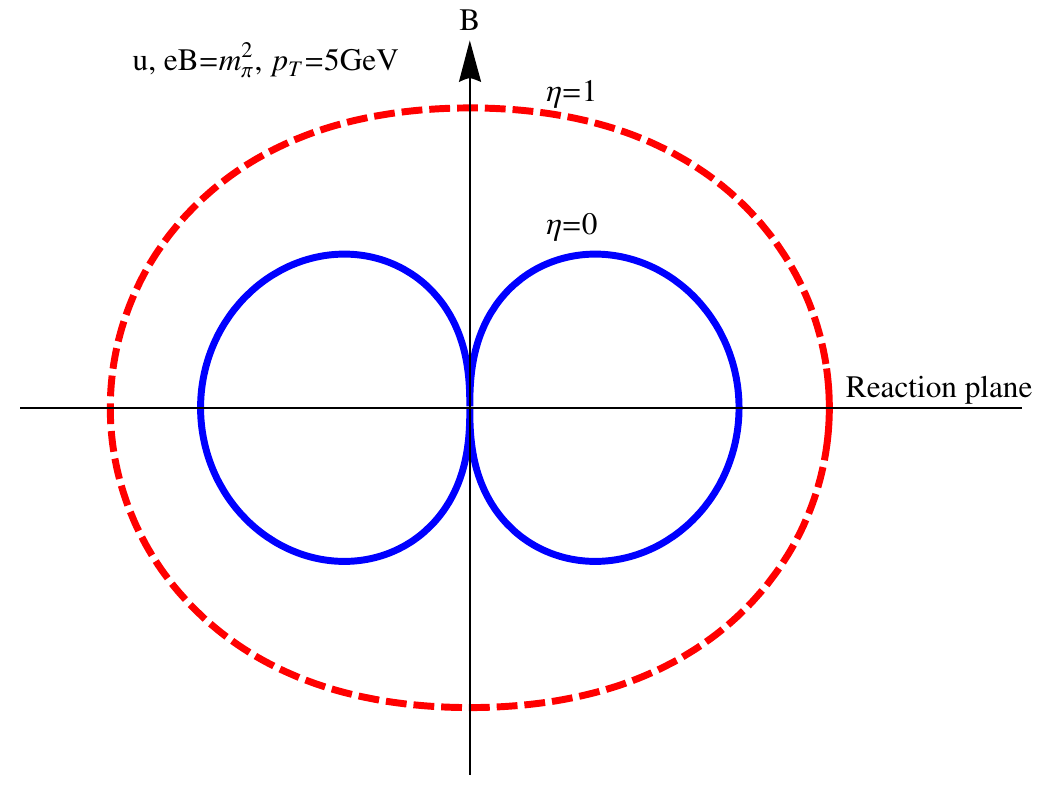} &
      \includegraphics[height=1.8in]{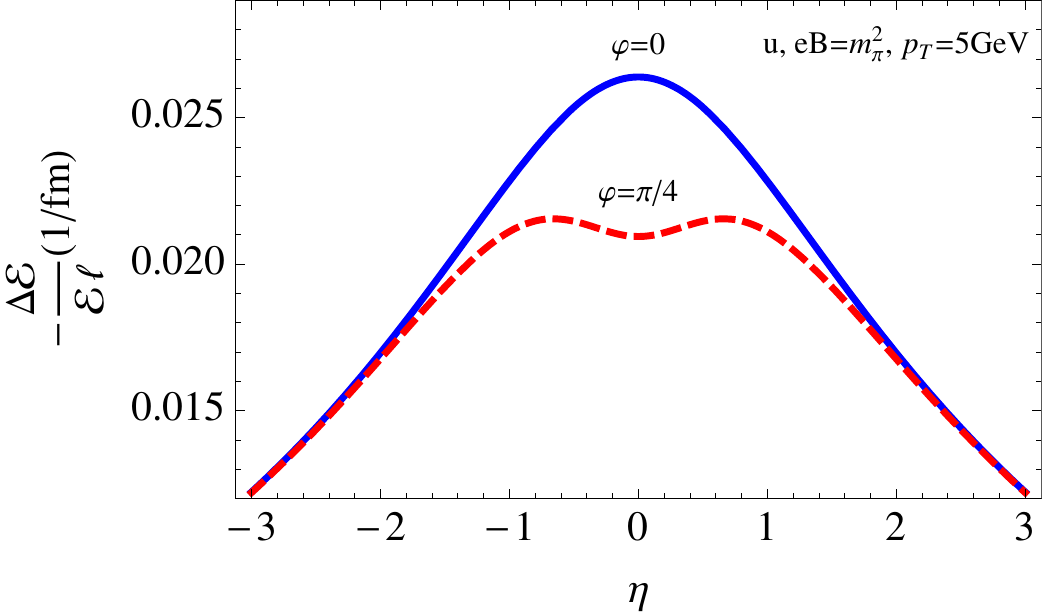}\\
        $(c)$ & $(d)$
      \end{tabular}
  \caption{Energy loss \emph{per unit length} by  quarks moving in constant external magnetic field  as a function of (a)  $p_T$ at RHIC at $\eta=\varphi=0$, (b)  $p_T$ at LHC at $\eta=\varphi=0$, (c)  azimuthal  angle $\varphi$ with respect to the reaction plane for $p_T=5$~GeV and $\eta=0,1$. (d)  Fractional energy loss vs rapidity $\eta$ at $p_T=5$~GeV and $\varphi= 0,\pi/4$.  }
\label{fig:loss-const}
\end{figure}

In \fig{fig:loss-const} we show the calculation of the energy loss \emph{per unit length} in a constant magnetic field using \eq{enegyloss} and \eq{chi-hi}. We see that energy loss  of a $u$ quark  with $p_\bot = 10$~GeV is about 0.2~GeV/fm at RHIC and 1.3~GeV/fm at LHC. This corresponds  to the loss of 10\% and 65\% of its initial energy  after traveling 5~fm at RHIC and LHC respectfully. Therefore, energy loss due to the synchrotron radiation at LHC gives a phenomenologically important contribution to the total energy loss. 

Energy loss due to the synchrotron radiation has a very non-trivial azimuthal angle and rapidity dependence that comes from the corresponding dependence of the $\chi$-parameter \eq{chi-hi}. As can be seen in  \fig{fig:loss-const}(c), energy loss has a minimum at 
$\varphi=\pi/2$ that corresponds to quark's transverse momentum $\v p_\bot$ being parallel (or anti-parallel) to the field direction. It has a maximum  at $\varphi= 0,\pi$ when $\v p_\bot$ is perpendicular to the field direction and thus lying in the reaction plane. It is obvious from \eq{chi-hi} that at mid-rapidity $\eta=0$ the azimuthal angle dependence is much stronger pronounced than in the forward/backward direction. 
Let me emphasize, that  the energy loss \eq{enegyloss} divided by $m^2$, i.e.\ $d\e/(dl\, m^2)$ scales with $\chi$. In turn, $\chi$ is a function of magnetic field, quark mass, rapidity and azimuthal angle. Therefore,  all the features seen in \fig{fig:loss-const} follow from this scaling behavior. 

\section{Polarization of light quarks }\label{sec:pol}

 Synchrotron radiation leads to polarization of   electrically charged fermions; this is  known as the Sokolov-Ternov effect \cite{Sokolov:1963zn}.  
Unlike energy loss that we discussed so far, this is a purely quantum effect. 
It arises because the probability of the spin-flip transition depends on the orientation of the quark spin with respect to the direction of the magnetic field and on the sign of fermion's electric charge. The spin-flip probability per unit time reads  
\cite{Sokolov:1963zn}
\beql{wst}
w=\frac{5\sqrt{3}\as C_F}{16}\frac{\hbar^2}{m^2}\left( \frac{\e}{m}\right)^5\,\left( \frac{Z_qe\,|\v v\times \v B|}{\e}\right)^3\,\left( 1-\frac{2}{9}\,(\v\zeta\cdot \v v)^2-\frac{8\sqrt{3}}{15}\mathrm{sign}\,(e_q)\,(\v \zeta\cdot \hat B)\right)\,,
\eeq
where 
$\v\zeta$ is a unit  axial vector that coincides with the direction of  quark spin in its rest frame, $\v v= \v p/\e$ is the initial fermion velocity. 

The nature of this spin flip transition is transparent in the non-relativistic case, where it  is induced by the interaction \cite{Jackson:1975qi}
\beql{nr}
H=-\v \mu\cdot \v B= -\left( \frac{geZ_q\hbar}{2m}\right)\,\vec s\cdot \vec B\,,
\eeq
It is seen, that negatively charged quarks and anti-quarks  (e.g.\ $\bar u$ and $ d$) tend to align against the field, while the positively charged ones (e.g.\ $ u$ and $\bar d$) align parallel to the field. 

Let $n_{\uparrow(\downarrow)}$ be the number of fermions or anti-fermions with given momentum and spin direction parallel (anti-parallel) to the field produced in a given event. At initial time $t=t_0$ the spin-asymmetry defined as 
\beql{as1}
 A = \frac{n_\uparrow-n_\downarrow}{n_\uparrow+n_\downarrow}\,
\eeq
vanishes. Eq.~\eq{wst} implies that at later times, a beam of  non-polarized  fermions develops a finite asymmetry   
given by \cite{Sokolov:1963zn}
\beql{as2}
A= \frac{8\sqrt{3}}{15}\left(1-e^{-\frac{t-t_0}{\tau}}\right)\,,
\eeq
 where
 \beql{tau}
 \tau= \frac{8}{5\sqrt{3}\, m\, \alpha_s C_F}\,\left(\frac{m}{\e}\right)^2\,\left( \frac{m^2}{Z_qe|\v v\times \v B|}\right)^3\,
\eeq
is the characteristic time (do not confuse with \eq{t-relax}!) over which the maximal possible asymmetry is achieved. This time is extremely small on the scale of $t_0$. For example, it takes only $\tau=6\cdot 10^{-8}$~fm for a $d$ quark of momentum $p_\bot =1$~GeV at $\eta=\varphi=0$ at RHIC to achieve the maximal asymmetry of $A_m= \frac{8}{5\sqrt{3}}=92$\%. Therefore, quarks and anti-quarks are polarized almost instantaneously after being released from their wave functions. Note, that  for quarks moving nearly parallel to the field direction, $\tau$ can be  large. However, the corresponding azimuthal angle and rapidity are always much smaller than their experimental resolution.  

As the quarks propagate through the hot medium their polarization is washed-out. This however does not seem to be an important effect because (i) the quark-gluon plasma is chirally symmetric phase of QCD and (ii) the spin-field interaction energy, estimated for, say, $d$ quarks from  \eq{nr} is of order 1.5~GeV, while the temperature in the plasma is of order of 0.2~GeV.
Nevertheless, we do expect that  the polarization of quarks is completely washed out in the fragmentation process at $T<T_c$  because the chiral symmetry is broken in the QCD vacuum. 

The Sokolov-Ternov effect can be observed by studying the polarization of light charged \emph{leptons}. The characteristic time $\tau$ is still very small. Indeed, in \eq{tau} one replaces $\as C_F$ by $\alpha_\mathrm{em}$ which -- in the case of electron -- is compensated by smaller mass.  In my opinion, observation of such a lepton polarization asymmetry would be the most definitive proof of existence of the strong magnetic field at  early times after a heavy-ion collision  regardless of its later time-dependence.

\section{Conclusions}\label{sec:dis}

In this paper we argued that the magnetic field created by fast heavy ions can be considered  as approximately constant due to high electric conductivity of the quark-gluon plasma.  We then used some well-known QED results to estimate  the energy loss suffered by fast quarks in external magnetic field in heavy-ion collisions. We found that the energy loss per unit length for a light quark with $p_T=15$~GeV is about 0.27~GeV/fm at RHIC and 1.7~GeV/fm at LHC, which is comparable to the losses due to interaction with the plasma. Thus, the synchrotron radiation alone is be able to account for quenching of jets at LHC with $p_\bot$ as large as 20~GeV. Synchrotron radiation seems to be one of the missing pieces in the puzzle of the jet energy loss in heavy ion collisions. 

We also pointed out that quarks and leptons are expected to be strongly polarized in plasma in the direction parallel or anti-parallel to the magnetic field depending on the sign of their electric charge. While polarization of quarks  is expected to be washed out during the fragmentation, that of leptons should survive to present a direct experimental evidence for the existence of the  magnetic field. Finally, we would like to mention a possible deep 
connection between  the ``tunneling through the horizon" thermalization mechanism that we discussed before \cite{Kharzeev:2005iz} and the polarization of fermions in external magnetic field that can be viewed as an Unruh effect \cite{Bell:1986ir,Leinaas:1998tu,Akhmedov:2006nd}.

\acknowledgments
I  am grateful to Dima Kharzeev for many informative discussions. I would like to thank Ajit M. Srivastava for pointing out a mistake in the estimate of  electrical conductivity in an earlier version of the paper.  
This work  was supported in part by the U.S. Department of Energy under Grant No.\ DE-FG02-87ER40371. I 
thank RIKEN, BNL, and the U.S. Department of Energy (Contract No.\ DE-AC02-98CH10886) for providing facilities essential
for the completion of this work.



\end{document}